\documentclass[prl,nofootinbib,amsmath,twocolumn,notitlepage,preprintnumbers]{revtex4-1}

\usepackage{graphicx,xfrac}  
\usepackage{dcolumn}   
\usepackage{bm}        
\usepackage{amssymb, amsmath}
\usepackage{slashed}   

 \newcommand{\brac}[2]{ \left( \frac{#1}{#2} \right) } 

\usepackage[usenames,dvipsnames,svgnames,table]{xcolor}

\def\be{\begin{eqnarray}}
\def\ee{\end{eqnarray}}
\def\bea{\begin{eqnarray}}
\def\eea{\end{eqnarray}}

\newcommand{\mpl}{M_{\rm Pl}}

\newcommand{\dneff}{\Delta N_{\rm eff}}

\usepackage{calrsfs}
\DeclareMathAlphabet{\pazocal}{OMS}{zplm}{m}{n}

\newcommand{\Eq}[1]{Eq.~(\ref{#1})}

\begin{document}

\preprint{FERMILAB-PUB-20-229-A-T}

\title{Dark Radiation from Inflationary Fluctuations } 

\author{Gordan Krnjaic$^{a,b}$}
\email{krnjaic@fnal.gov}
\thanks{ORCID: http://orcid.org/0000-0001-7420-9577}

\affiliation{$^a$Fermi National Accelerator Laboratory, Theoretical Physics Department}
\affiliation{$^b$University of Chicago, Kavli Institute for Cosmological Physics}

\date{\today}




\begin{abstract}
Light new vector bosons  
can be produced gravitationally through quantum
fluctuations during inflation; if these particles are feebly coupled and cosmologically metastable, they 
can account for the observed dark matter abundance. However,
in minimal anomaly free $U(1)$ extensions to the Standard Model, 
these vectors generically decay to neutrinos if at least one neutrino mass eigenstate is sufficiently light. If these decays occur between neutrino decoupling and CMB freeze out, the resulting radiation energy
density can contribute to $\dneff$ at levels that can ameliorate the Hubble tension and 
be discovered with future CMB and relic neutrino detection experiments. Since the additional neutrinos are 
produced from vector decays after BBN, this scenario predicts
 $\Delta N_{\rm eff} > 0$ at recombination, but $\dneff = 0$ during BBN.  Furthermore, due to a fortuitous cancellation, the contribution to $\dneff$ is approximately mass independent.

\end{abstract}

\maketitle




\section{Introduction}

Cosmological inflation elegantly accounts for the observed flatness, isotropy, and homogeneity
of the universe. Additionally, the quantum mechanical fluctuations in the inflaton field during inflation
 generate a nearly scale invariant spectrum of 
density perturbations that seed 
the growth of structure and 
imprint temperature anisotropies onto the cosmic microwave background (CMB) -- see Ref. \cite{baumann2009tasi} for a review.

It is well known that new, feebly coupled particles are produced gravitationally through quantum 
fluctuations during inflation if their masses are small compared to the inflationary Hubble scale $H_I$ \cite{Mukhanov:1990me}; heavier particles can also be produced if the inflaton undergoes rapid oscillations
\cite{Chung:1998zb,Kuzmin:1998kk,Ema:2015dka,Ema:2016hlw,Ema:2018ucl,Chung:2018ayg} or nontrvially affects the particle's mass during inflation \cite{Fedderke:2014ura}.
  For light spin-0 particles, these fluctuations yield isocurvature perturbations on large scales, which
are tightly constrained by CMB observations \cite{PhysRevD.32.3178,Linde:1985yf} and for spin $\sfrac{1}{2}$ fermions, inflationary fluctuations are generically suppressed unless they have non-conformal interactions through higher dimension operators \cite{Parker:1968mv,Adshead:2015kza,Adshead:2018oaa}.

It has recently been shown that the gravitational production of spin-1
 particles during inflation is sharply peaked at modes that re-enter the horizon 
 after inflation when the Hubble scale equals the vector's mass, $H = m$ \cite{Graham:2015rva}. Such scales are typically 
 much smaller than those probed by CMB experiments, so the isocurvature bounds 
 on this scenario are negligible and this mechanism yields a viable
 dark matter candidate for 
\be
m \sim \mu {\rm eV} \left( \frac{10^{14} \, \rm GeV}{ H_I} \right)^{4} ~.
\ee
Thus, if the vector is decoupled from Standard Model (SM) fields or is  sufficiently light ($m \ll 2 m_e$) and interacts only through a small kinetic mixing, 
its cosmological metastability is generically realized.\footnote{For a kinetically mixed $V$, allowed 
decays $V\to 3\gamma$ are highly suppressed \cite{Pospelov:2008jk,McDermott:2017qcg} and if
the vector kinetically mixes with SM hypercharge before electroweak symmetry breaking, 
decays to $V \to \bar \nu \nu$ are further suppressed by powers of $\sim (m/m_Z)^4$ \cite{Hoenig:2014dsa}.}
 
However, if the vector is the gauge boson of a minimal $U(1)$ gauge extension, couplings to neutrinos are  required for anomaly cancellation \cite{Bauer:2018onh}; the only anomaly free groups with 
no additional SM charged fermions are 
\be
\label{uone}
~~U(1)_{B-L}, ~~U(1)_{L_i-L_j},~~ U(1)_{B-3L_i},
\ee
 where $B/L$ is baryon/lepton number, $i,j = e,\mu,\tau$ are lepton flavor indices, and the corresponding
 gauge bosons in these models couple to at least one neutrino flavor.
Thus, unlike  kinetically mixed dark photon scenarios, the 
vector decays in these models can be relatively prompt and have observable cosmological
consequences. 

In this paper, we consider the fate of light gauge bosons $V$ produced during inflation.
We assume these vectors couple feebly to neutrinos and that at least one neutrino mass eigenstate is sufficiently
light to allow $V \to \bar \nu \nu$ decays. If such decays occur after neutrino decoupling, but before
CMB photon decoupling, there is an irreducible contribution to $\Delta N_{\rm eff}$ 
 that is potentially observable
with future CMB-S4 experiments \cite{Abazajian:2016yjj} and a modified relic neutrino spectrum
observable at PTOLEMY  \cite{McKeen:2018xyz,Baracchini:2018wwj}. Furthermore, such a 
contribution of $\dneff$ can alleviate the discrepancy between early and late time measurements of the Hubble constant 
(for recent reviews see \cite{divalentino2021realm,Knox_2020}).




\section{Stable Vector Abundance}

The general lagrangian during inflation contains
 \be
 \label{lag}
\frac{ \cal L}{\sqrt{\tilde g}} \supset      -\frac{1}{4} g^{\mu\kappa}g^{\nu\lambda} F_{\mu\nu}F_{\kappa \lambda} + \frac{m^2}{2} g^{\mu\nu} V_\mu V_\nu ~,
 \ee
where $V$ is a gauge boson in an FRW metric, $F_{\mu\nu}$ is the corresponding
field strength tensor, and $\tilde g$ is the metric determinant. 
If the mass satisfies $0 < m \ll H_I$ and $V$ is stable, the longitudinal mode\footnote{The transverse mode is conformally coupled to gravity, so its production
is greatly suppressed by comparison \cite{Graham:2015rva}.}
is gravitationally produced  during inflation and constitutes a present-day dark matter fraction 
$f^0_{V} \equiv \Omega_{V}/\Omega_{\rm dm}$  \cite{Graham:2015rva}
\be
\label{f0}
f^0_{V} \approx
\frac{\sqrt{m} H_I^2}{4\pi^2 M^{3/2}_{\rm Pl} T_{\rm eq}} \approx
   10^{-2} \sqrt{ \! \frac{m}{10 \,\mu\rm eV} } \left(  \frac{H_I}{10^{13} \rm \, GeV} \right)^2 \!\!\!,~~~
\ee
where $\mpl = 1.22\times 10^{19}$ GeV is the Planck mass and $T_{\rm eq} = 0.75$ eV is the 
temperature of matter-radiation equality, so the energy density at earlier times is
\be
\label{rhoV}
\rho_V(t) = \rho_V^0 \left( \frac{a(t_0)}{a(t)} \right)^3,~~ 
\rho_V^0 \equiv f^0_V \, \Omega_{\rm dm} \, \rho_{\rm cr}~,
\ee
where  $\Omega_{\rm dm} = 0.24$ is the fractional dark matter abundance, $\rho_{\rm cr} = 4.1 \times 10^{-47} \, \rm GeV^4$ is the critical density,  $a$ is the FRW scale factor, $t_0 = 13.8$ Gyr, 
and a $0$ label represents a present day quantity \cite{Aghanim:2018eyx,Kolb:1990vq}. 
For stable vectors, \Eq{rhoV} is valid for $t > t_\star = (2m)^{-1}$, the horizon re-entry time 
corresponding to $H = m$ and temperature 
\be
\label{Tstar}
T_\star = \sqrt{    \frac{ m M_{\rm Pl}}{1.66 \, \sqrt{g_\star} }       }
 \approx 85 \, {\rm GeV} 
 \left( \frac{100}{g_\star}\right)^{1/4}
 \!\!
 \left( \frac{m}{10  \, \mu \rm eV}\right)^{1/2} \!\!\!\!\!\!,~~~
\ee 
where $g_\star$ is the effective number of relativistic SM species in equilibrium. 
Note that because the $V$ power spectrum is dominated by 
momentum modes that re-enter the horizon when $H\sim m$,
the $V$ population is nonrelativistic for all times $t > t_\star$. 

 


\section{Adding Decays to Neutrinos}
\label{decays}

Since abelian gauge extensions to the SM generically feature 
neutrino couplings, we add the representative interaction 
\be
{\cal L} \supset  g V_\mu \bar \nu_i \gamma^\mu \nu_i~,
\ee
to \Eq{lag}, where $g \ll 1$ is a gauge coupling and $i$ is a lepton family index. 
In the massless neutrino limit, the partial width to a single flavor is 
 \cite{Escudero:2019gzq}
\be
\Gamma(V \to \bar \nu_i \nu_i) = \frac{g^2 m}{24 \pi}~,
\ee
 the total width $\Gamma_V$ is the sum of all allowed channels and
 $\tau_V= \Gamma_V^{-1}$ is the $V$ lifetime.
We note that a single massless neutrino eigenstate is empirically viable \cite{Tanabashi:2018oca,deSalas:2018bym},
so, in principle, at least one decay channel is allowed for all vector masses. 

Unlike in Ref \cite{Graham:2015rva}, here the vector is unstable and $V\to \bar \nu \nu$ decays deplete the
initial population, so \Eq{rhoV} is only useful for establishing the initial condition for $\rho_V$ at $t=t_\star$.
Accounting for decays to neutrinos, the $V$ population can now be written
\be
\label{rho-t}
\rho_V(t) = \rho^0_V \left(  \frac{ a(t_0)}{a(t)} \right)^3 e^{-\Gamma_V (t- t_\star)}~,
\ee    
and the energy density of the modified neutrino population $\delta \rho_\nu$ 
evolves according to 
\be
\label{diffeq}
 \delta \dot \rho_\nu + 4 H \delta\rho_\nu = \Gamma_V \rho_V~,
\ee
which can be integrated to yield 
\be
\label{delnu}
\delta \rho_\nu(t) =  \frac{\Gamma_V}{a(t)^4} \int_{t_\nu}^{t} dt^\prime a(t^\prime)^4\rho_V(t^\prime) ~,
\ee
where $a$ is the FRW scale 
factor and $t_\nu \sim$ 1 sec is the time of neutrino decoupling;
 we only keep contributions for $t > t_\nu$ because neutrinos injected
before $t_\nu$ thermalize with the radiation bath and do not contribute to dark radiation. 
Similarly, $V$ that decay after CMB decoupling will not contribute to $\dneff$, but will
increase the dark matter density during recombination. In Fig. \ref{densities} we show a 
representative solution of \Eq{diffeq} plotted as a fraction of the total energy density. 
 
  In terms of the equivalent number of SM neutrinos  $\dneff$, this additional radiation from $\delta \rho_\nu$
  predicts 
\be
\label{neff}
\dneff \equiv \frac{8}{7}\left(  \frac{11}{4} \right)^{4/3} \frac{\delta \rho_{\nu}}{\rho_\gamma}  \, \biggr |_{T_{\rm cmb}},
\ee
where $\rho_\gamma = \pi^2 T^4/15$ 
and the contribution is evaluated at the temperature of photon decoupling, $T_{\rm cmb} \approx 0.2$ eV;
this sets the upper integration range in \Eq{delnu} since $V$ decays after last scattering do not contribute to 
dark radiation in the CMB data set. 

\begin{figure}
\hspace{-0.2in}
\includegraphics[width=3.3in,angle=0]{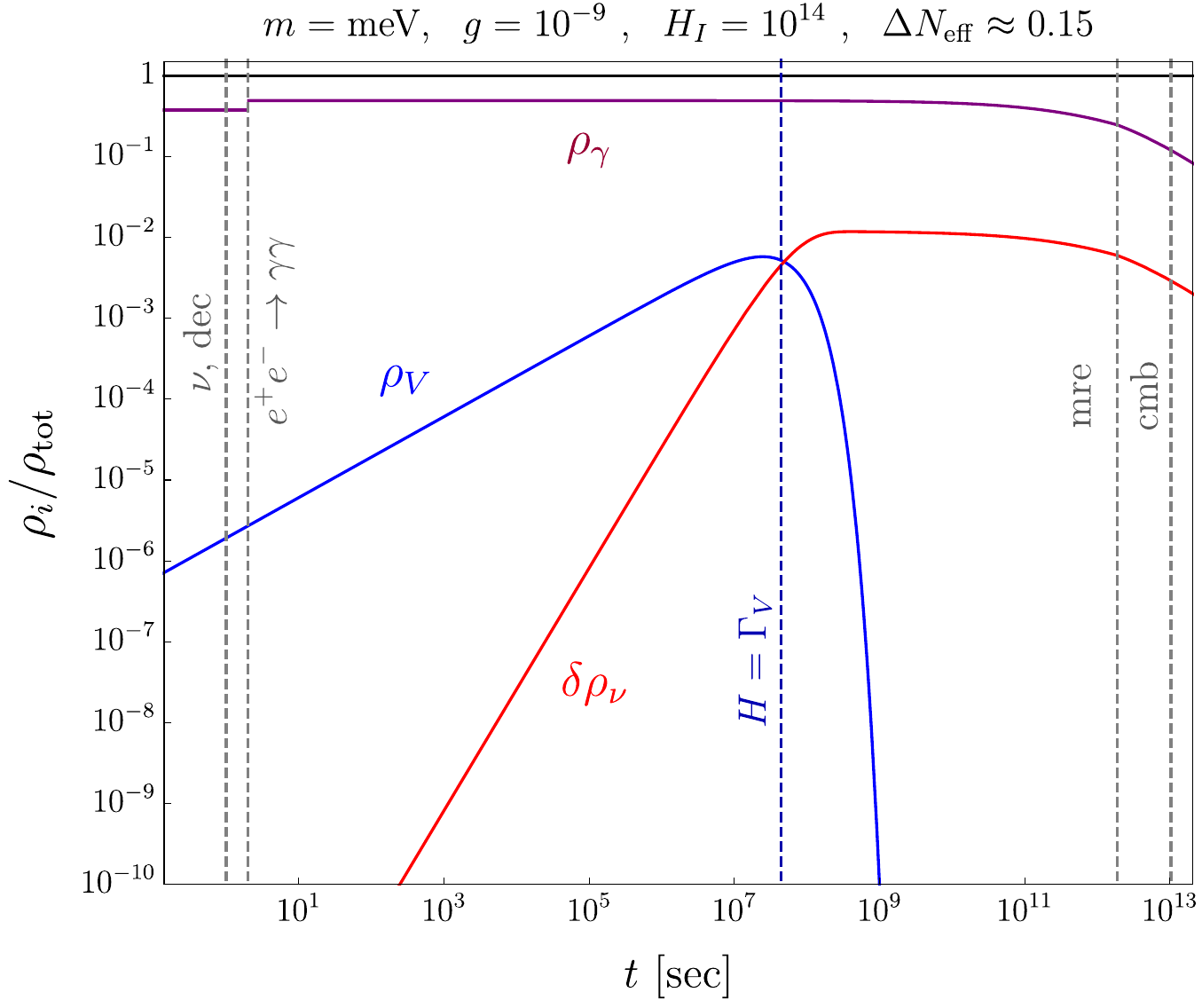}
\caption{ 
Time dependent energy fractions  $\rho_i /\rho_{\rm tot}$ for a benchmark choice of model inputs.
Here $\rho_{\rm tot} = 3 M_{\rm Pl}^2 H^2/ 8\pi$ is the total energy density of the universe and 
we show $\rho_V$, the density of vectors from inflationary production, 
$\delta \rho_\nu$ the additional neutrino density from $V\to \bar \nu \nu$ decays assuming a single neutrino flavor. From left to right, the vertical
dashed lines mark neutrino decoupling, matter radiation equality, and CMB decoupling.
Note that in \Eq{dave}, the number density of the  neutrino population from $V$ decays might exceed 
that of the relic neutrino background, but as seen here, the energy density remains small
for empirically viable values of $\Delta N_{\rm eff}$.
 }
\label{densities}
\end{figure}




\begin{figure}
\hspace{-0.0in}
\includegraphics[width=3.3in,angle=0]{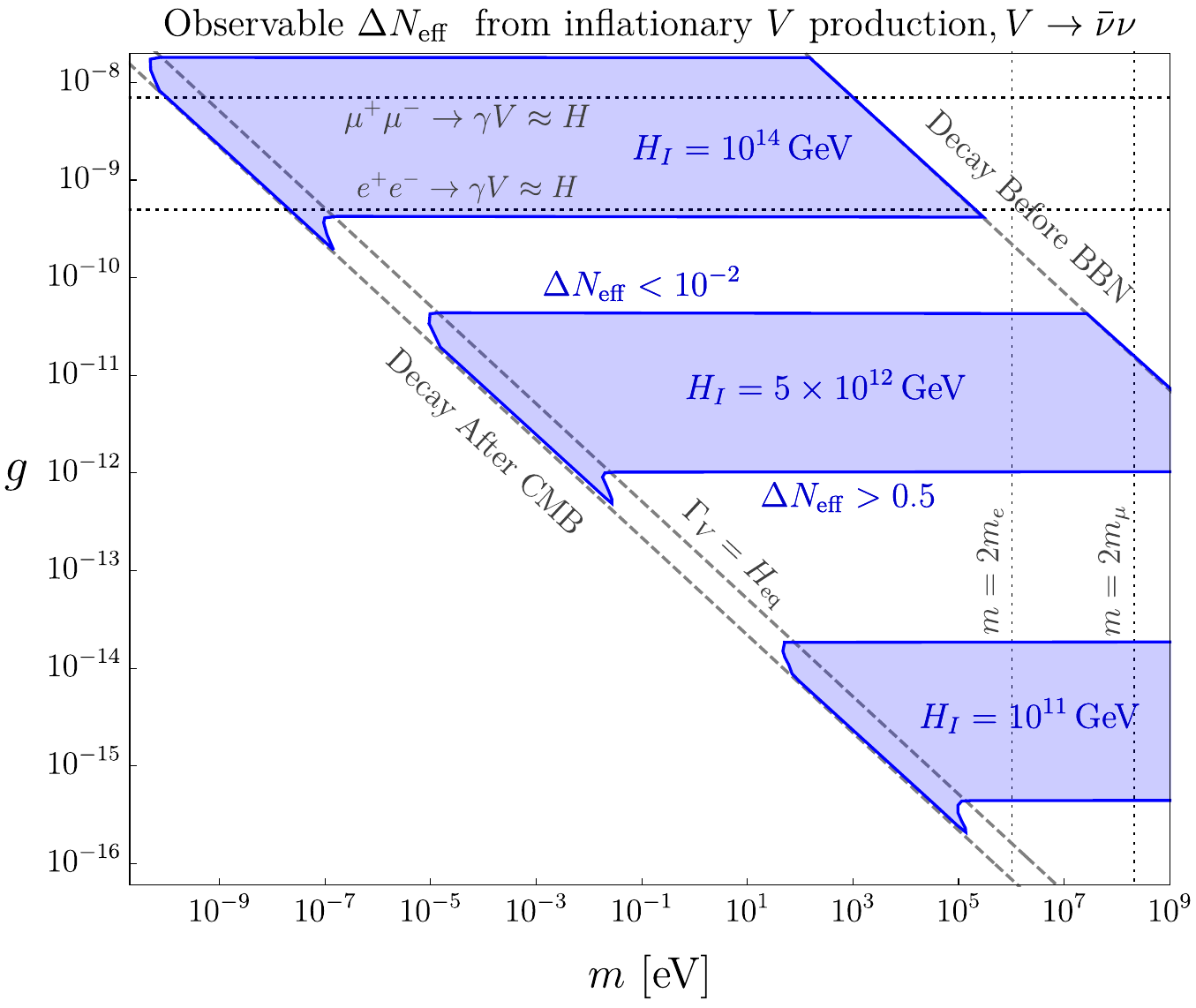}
\caption{ Parameter space that yields observable levels of dark radiation from a
 population of gravitationally produced vectors that decay via $\to \bar \nu \nu$
 after neutrino decoupling but before recombination.
Horizontal blue shaded bands represents  regions where
  $10^{-2} <\dneff < 0.5$ for representative choices
  of the inflationary Hubble scale $H_I$; for each choice, the parameter
  space below the bottom boundary predicts $\Delta N_{\rm eff} > 0.5$,
  which is excluded assuming otherwise standard cosmological assumptions
   \cite{Blinov:2019gcj,Depta:2019lbe,Aghanim:2018eyx}. Above the   horizontal dotted lines, $V$ thermalizes with the SM, yielding
   $\dneff \approx 2.5$ \cite{Blinov:2019gcj}, which is excluded
   if $V$ couples to $e$ or $\mu$. The vertical dotted lines 
   mark $m = 2m_{e,\mu}$ where 
 $V \to e^+e^-$ and   $V \to \mu^+\mu^-$
    decays are kinematically allowed. Most models
    in \Eq{uone} feature $V$-$e$ couplings, so for $m > 2m_e$ 
    the $\dneff \approx 0$ as $V \to e^+e^-$ decays heat photons to 
    compensate for $V \to \bar \nu\nu$ decays, which heat neutrinos.
 }
\label{direct}
\end{figure}

For the full parameter space, $\dneff$ in \Eq{neff} must be computed numerically  
by solving \Eq{delnu}. However, if $V$ decays between $T_{\nu, \rm dec}$ and $T_{\rm eq}$, 
the decay temperature can be written
\be
T_{\rm decay} \approx
 \sqrt{   \frac{ g^2 m M_{\rm Pl}}{40 \pi \sqrt{g_\star}}      } 
 \approx 100 \, {\rm eV} \! \left( \frac{g}{10^{-8} }\right) \! \left( \frac{m}{10^{-5} \, \rm eV}\right)^{1/2} \!\!\!\!\! ,~~~~~~
\ee
where $g_\star \approx 3.36$ in our temperature range of interest between decoupling and recombination. 
Assuming instantaneous $V \to \bar \nu \nu$ decay and
 approximating $\delta \rho_\nu \approx \rho_V(T_{\rm decay})$ using \Eq{rhoV},
  \Eq{neff} becomes
\be
\dneff \label{dneff-rad}
&\approx &
 \frac{30}{7 \pi^4} \left( \frac{11}{4} \right)^{4/3}
\frac{ \Omega_{\rm dm}  \,
\rho_{\rm cr}
\sqrt{m} H_I^2}{ M^{3/2}_{\rm Pl}   T_0^3  T_{\rm eq} T_{\rm decay} } \nonumber \\
 &\approx&
10^{-2}  \left( \frac{H_I}{10^{14}  \, \rm GeV} \right)^2
\left( \frac{10^{-8}}{g}  \right)~,~~
\ee
where the vector mass has canceled.

 In Fig. \ref{direct} we show
$\dneff$ predictions for the inflationary vector population where 
we compute $\delta \rho_\nu$ numerically using \Eq{delnu}. The 
blue horizontal bands represent the currently viable $10^{-2} \le \dneff < 0.5$
range that is within the reach of  CMB-S4 predictions \cite{Abazajian:2016yjj}.
Note that current BBN bound $\dneff < 0.5$ \cite{Blinov:2019gcj} is less stringent than
 the  CMB and large scale structure bound $\dneff < 0.28$ \cite{Aghanim:2018eyx}, but 
 the BBN limit is less model dependent because it is not as sensitive to the choice
of cosmological model. However, despite the nominal choice of $\dneff < 0.5$ as our
conservative exclusion benchmark, this scenario is not directly constrained by
the BBN measurement of $\dneff$ since the additional neutrinos
from $V$ decays do not appear until after BBN. 

The area in between the dashed diagonal bands represent parameter space  
for which $V\to \bar \nu \nu$ decays occur between neutrino and CMB decoupling; 
decays outside this band do not contribute to $\dneff.$ The vertical lines
at $m = 2 m_e, 2m_\mu$ represent regions where the $\dneff$ prediction here
does not apply if $V$ couples to electrons or muons; in such models, $V$ decays
to charged particles after neutrino decoupling will heat photons and thereby 
reduce $\dneff$ relative to \Eq{dneff-rad}.

We note for completeness that there is also a possible contribution to $\dneff$ from
the $V$ population itself if an appreciable fraction of the $\rho_V$ redshifts like radiation
at recombination. Since inflationary $V$ production is sharply
peaked around modes that enter the horizon at $H \sim m$, from \Eq{Tstar} only masses below $m \lesssim 10^{-30}$ eV
will be quasi relativistic around $T_{\rm cmb}$. However, from \Eq{f0} such small masses 
yield negligible inflationary production for all $H_I \lesssim 10^{14}$ GeV allowed by CMB limits 
on tensor modes \cite{Aghanim:2018eyx,Marsh_2016}, so we can safely neglect this contribution.




\section{Interactions with the SM Plasma}

The above discussion assumes that the early universe $V$ population 
arises entirely to  inflationary production and 
is unaffected by the SM radiation bath. However,
for any value of the gauge coupling, there is irreducible
sub-Hubble ``freeze-in" production of additional $V$
\cite{Dodelson:1993je,Hall:2009bx,Fradette:2014sza,Escudero:2019gzq} 
and, if the coupling is sufficiently large, the $V$ population can
thermalize with the SM plasma; which yields additional contributions to $\dneff$.

\begin{itemize}
\item{\bf Inverse Decays}
\\
Independently of any other assumptions about ultralight $V$ partilces beyond their coupling 
to neutrinos, there is a bound on thermalizing with the SM plasma via
population via $\bar \nu \nu \leftrightarrow V$ decays and inverse decays. 
If thermalization occurs before neutrino decoupling,
this scenario predicts $\dneff \approx 2.5$, so avoiding this fate requires
\be
~~~~~~\frac{\Gamma_{\bar \nu \nu \to V}}{H} \!  \sim \! \frac{g^2 m^2 \mpl}{T_{\nu, \rm dec}^3} \! \ll  \!  1     \! \implies\!
 g \lesssim 10^{-5} \!  \left( \frac{\rm eV}{m} \right) \! ,~~~
\ee
where $T_{\nu, \rm dec} \sim $ MeV is the temperature of neutrino decoupling via the
SM weak interactions. If, instead, thermalization occurs between $T_{\nu,\rm dec}$ and 
$T_{\rm cmb}$ as in Ref. \cite{Berlin:2017ftj}, then $\dneff \sim 0.2$
independently of mass and coupling \cite{Escudero:2019gzq}.\footnote{Although
Ref. \cite{Escudero:2019gzq} specifically considered the gauged $L_\mu -L_\tau$ scenario,
this conclusion holds for any ultralight vector $m \ll m_e$ with a coupling
to neutrinos, which includes all anomaly free $U(1)$ extensions 
that gauge global SM quantum numbers \cite{Bauer:2018onh} }  Since this contribution is
fixed only by the neutrino coupling, it must be added to the component from the
inflationary population. 

\item{\bf Production From Charged Particles} 
\\
If $V$ also couples to charged fermion $f$, 
dangerous $\bar f f \to \gamma V$  
and  $f\gamma \to f V$ processes can thermalize $V$ with 
the SM radiation bath, thereby yielding $\dneff \approx 2.5$, which
is excluded by both BBN and CMB observables \cite{Escudero:2019gzq,Blinov:2019gcj,Depta:2019lbe,Aghanim:2018eyx}.\footnote{This $\dneff \approx 2.5$ prediction assumes that the
the thermalized $V$  population does not decay before neutrino decoupling,
which is true for the  entire parameter space we consider here.}
The $V$ production rate can be estimated as
$\Gamma_{\bar ff \to V \gamma } \sim \Gamma_{f \gamma \to f V } \sim \alpha g^2 T/4\pi$,
so these processes grow relative to Hubble until $T \sim m_f$, when
 they become Boltzmann suppressed. Ensuring that the maximum rate not exceed Hubble 
 expansion requires 
 \be
 \label{therm}
 g \lesssim \sqrt{ \frac{4\pi  \sqrt{g_\star} m_f}{\alpha M_{\rm Pl}} } =
 \begin{cases}
  5\times 10^{-10} ~,& f = e \\
  7\times 10^{-9} ~,& f = \mu
 \end{cases},
 \ee
 where $g_\star$ is evaluated at $T = m_e, m_\mu$, respectively.
The stronger electron based bound here applies to most anomaly
free $U(1)$ extensions -- including gauged $B-L$, $B- 3 L_e$,
 $L_e-L_\mu$, $L_e-L_\tau$ -- as they all require  $V$ to couple to
 electrons for anomaly cancellation \cite{Bauer:2018onh}; the main outlier is gauged
 $L_\mu -L_\tau$ for which muon induced thermalization is the 
 dominant process at low temperatures \cite{Escudero:2019gzq}, so the
 bound is somewhat weaker.  Both of the requirements in  \Eq{therm} are presented in 
as dotted horizontal  black curves in Fig. \ref{direct} and the parameter space above these
regions is excluded if the model in question features the corresponding
 $e$ or $\mu$ coupling.

\end{itemize}




\begin{figure}
\hspace{-0.in}
\includegraphics[width=3.3in,angle=0]{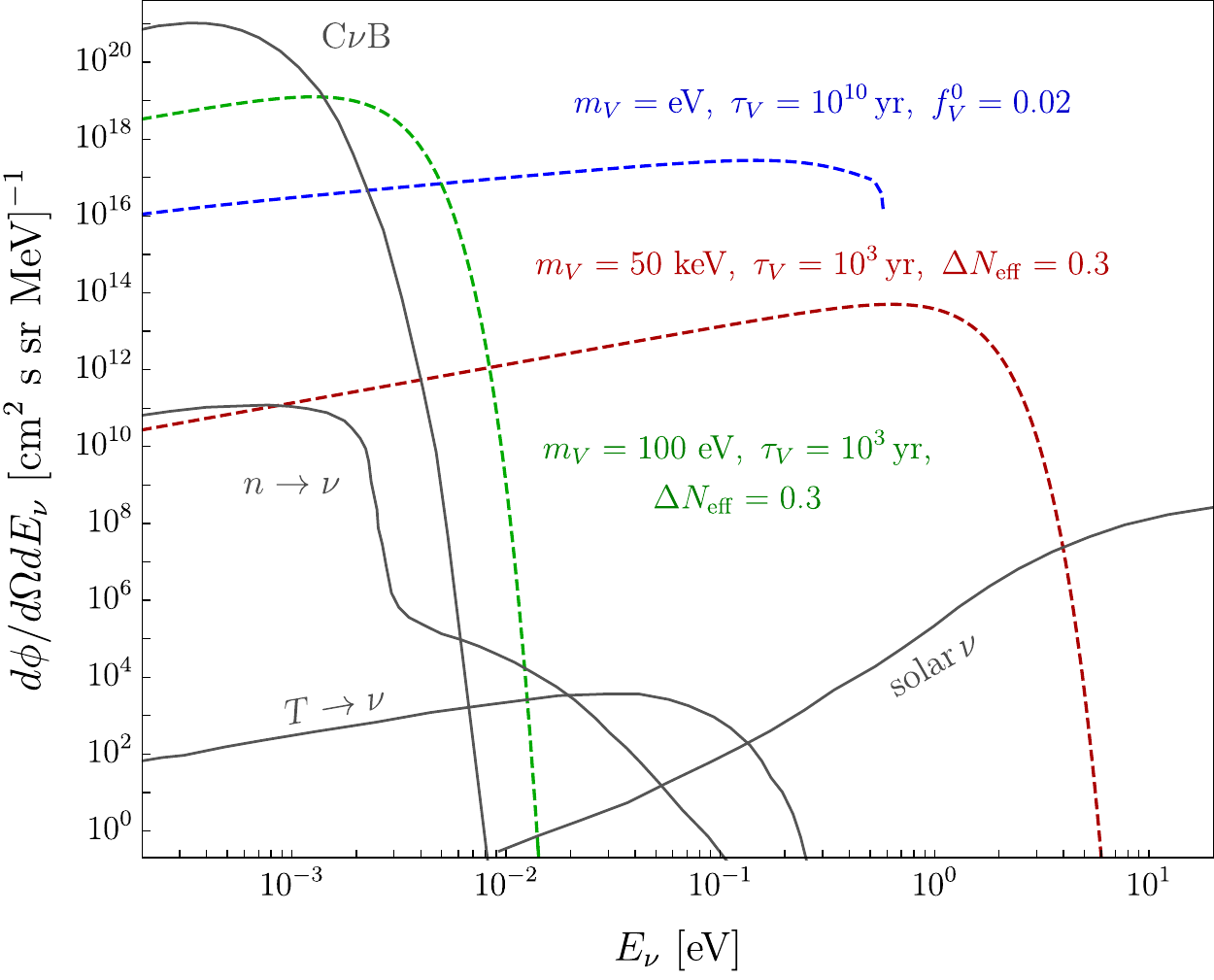}
\caption{ 
Present day neutrino flux spectra from $V \to \bar \nu \nu$ decays for representative
benchmark points (dashed). Also shown are spectra from
the C$\nu$B, primordial neutron decays
during BBN ($n\to \nu$), tritium decays during BBN $(T\to \nu)$,
and solar neutrinos \cite{McKeen:2018xyz}. 
From \Eq{rate-neff}, is clear that the early decaying parameter points (green and red) 
only yield appreciable ($\gtrsim$ few) events at PTOLEMY for lower values of $m_V$, 
which are difficult to distinguish from the C$\nu$B spectrum, but might
be detected as an enhancement over the Standard Model signal rate. 
As the mass is increased, the spectrum gets harder, but  the rate becomes 
unobservable with a feasible exposure; for the 50 keV benchmark, we find
$R\sim 10^{-3}$ events/yr at PTOLEMY. For later decaying particles (blue dashed curve)
the rate and spectrum can be favorable, but there is no contribution to $\dneff.$ 
 }
 \label{fluxfig}
\label{densities}
\end{figure}

We emphasize that the parameter space shown in Fig. \ref{direct} 
is extremely weakly coupled, such that there is no danger of 
the inflationary $V$ population thermalizing
with the SM plasma, or of any appreciable contributions
from SM processes that produce additional $V$ 
particles via inverse decays or SM scattering reactions. 
For a careful study of such processes in the context of the 
models studied here, see \cite{Escudero:2019gzq} which
identifies the parameter space where freeze in production
via inverse decays contributes to cosmological 
observables including $\Delta N_{\rm eff}$.




\section{Present Day Neutrino Flux}

In this section we review the results of Ref. \cite{McKeen:2018xyz}, 
adapted to the case of inflationary vector production.
The neutrinos in our scenario arises from
$V$ decays and if the entire population decays in the early universe,
the present day number density is   
\be 
\label{dave0}
\delta n_\nu(t_0) =  \frac{ f_V^0 \, \Omega_{\rm dm} \, \rho_{\rm cr}}{m_V} \approx  130 \, {\rm cm^{-3} } \, \brac{f_V^0}{0.05} \brac{\rm eV}{m}\!.~~~
\ee
If these decays occur between neutrino decoupling and recombination, 
 \Eq{dave0} can be rewritten \cite{McKeen:2018xyz}
\be 
\label{dave}
\delta n_\nu(t_0) \approx  10^3 {\rm cm}^{-3} \left(\frac{\dneff}{0.28} \right)
 \left(\frac{\rm eV}{m} \right) 
  \sqrt{   \frac{10^{3} \, \rm yr }{\tau_V} }.~~~~
\ee
Although the {\it number} density of additional
neutrinos in \Eq{dave} can exceed the $\sim 
300 {\rm \, cm}^{-3}$ number density of the cosmic neutrino background (C$\nu$B) as predicted
in the Standard Model, as long
as the corresponding value of $\Delta N_{\rm eff}$ satisfies observational bounds, the
energy density of this population is always subdominant and remains empirically viable.

For some parameter choices, this additional neutrino population may be observable with the
PTOLEMY experiment using inverse beta decay reactions
from captured relic neutrinos \cite{Baracchini:2018wwj}. 
Assuming a detector target mass of $M_T $, electron
neutrino fraction $f_{\nu_e}$, neutrino capture cross section 
on tritium $\sigma = 3.83 \times 10^{-45}$ cm$^{2}$, and the excess
neutrino density from \Eq{dave}, the signal rate is estimated to be 
\cite{McKeen:2018xyz}
\be
R   \approx \frac{5}{\rm yr} \brac{M_T}{100\,\rm g} \brac{f_{\nu_e} }{0.5} 
\brac{f_V^0}{0.05} \brac{\rm eV}{m_V}~,
\ee
which only assumes that the $V$ decay after decoupling. However,  
for $V$ that also decay before recombination, the fraction satisfies
\cite{McKeen:2018xyz}
\be
f_V^0 \approx 0.42 \brac{\dneff}{0.3} \brac{10^3 \, \rm yr}{\tau_V},
\ee
so the rate for early decaying $V$ can be written 
\be
\label{rate-neff}
R   \approx \frac{10}{\rm yr} \brac{M_T}{100\,\rm g} \brac{f_{\nu_e} }{0.5} 
\brac{\dneff}{0.3} \brac{\rm 10\, eV}{m_V}~,
\ee
which may be detectable with a year of exposure at PTOLEMY, 
whose projected C$\nu$B sensitivity is at the $\sim 10$ event level.
Note that there is general tension between having an appreciable $\dneff$ 
signal and having a distinguishable neutrino spectrum with a detectable
PTOLEMY rate. 

To see this, note that the late time flux of neutrinos from $V$ decays is
\be
\label{flux}
\frac{d\phi}{d\Omega dE_\nu} = \frac{f_V^0 \, \Omega_{\rm dm } \, \rho_{\rm cr}     }{ 2\pi m_V E_\nu}
\frac{ \Gamma_V\, e^{- \Gamma_V(t-t_\star)}}{H(z_c)}  \Theta(t-t_{\nu,\rm dec}),~~~
\ee
where $E_\nu$ is the 
observed energy of a present day neutrino emitted at redshift $z$ with
 energy $E_\nu(1+z)  = m_V/2$,
  $H(z) = H_0\sqrt{\Omega_\Lambda + \Omega_m(1+z)^3 + \Omega_r (1+z)^4}$ 
is the Hubble rate, $H_0  = 67$ km/sec/Mpc \cite{Aghanim:2018eyx}, and $z_c = [m_V/(2E_\nu)]-1$.
The theta function ensures that decays before neutrino decoupling do not contribute to the flux;
this population will thermalize with C$\nu$B. In Fig. \ref{fluxfig} we show 
representative flux spectra for both early ($t_{\nu,\rm dec} < \tau_V < t_{\rm cmb})$ and late time $(\tau_V > t_{\rm cmb})$ decaying 
populations. From \Eq{rate-neff}, early decaying $V$ with low $\sim$ few eV masses can yield appreciable
PTOLEMY signal rates, but as we see in these spectra, the fluxes similar to the 
C$\nu$ unless $m_V$ much greater, which trades off against the overall rate as $R\propto m_V^{-1}$.
It is possible to get an appreciable PTOLEMY flux for a low mass particle, but 
obtaining a distinctive spectral shape requires late time decays, which do not
affect $\dneff$.




\section{Conclusion}

In this paper we have studied the fate of massive vector particles produced gravitationally from inflationary fluctuations. If these vectors only interact with the SM via kinetic mixing, for $m<2 m_e$, the only allowed decay is 
$V \to 3 \gamma$ which is sharply suppressed, so $V$ is generically metastable
 can serve dark matter candidate \cite{Graham:2015rva}. 
However, if the vector arises in 
well motivated, minimal $U(1)$ gauge extensions from \Eq{uone}, it must couple to neutrinos, so if at least one neutrino mass eigenstate is sufficiently light, $V \to \bar \nu \nu$ decays can efficiently deplete this inflationary population and increase the relic neutrino densty, thereby predicting $\dneff \ne 0$. For certain regions of parameter space,
the same neutrino population may be observable at late times with the PTOLEMY experiment; for long lived vectors that decay after recombination, it is also possible to obtain an appreciable PTOLEMY signal even though
 $\dneff = 0$. 

Intriguingly. due to a cancellation, this contribution depends only  on $H_I$ and  $g$ as long as the $V$ lifetime falls within this time window. For a wide range of model
parameters, the $\dneff$ prediction in these scenarios is within reach of CMB-S4 projections \cite{Abazajian:2016yjj}. 
We note that, outside of the narrow parameter region where  50 keV $\lesssim T_{\rm decay} \lesssim $MeV, 
 this scenario predicts $\dneff \ne 0$ only in 
CMB data because nearly all of the $V$ decays occur after BBN has completed; decays before BBN 
thermalize with the SM, so $T_\nu/T_\gamma$ does not deviate from the SM prediction. However,
for parameter space in this decay occurs after recombination, the resulting neutrino population
may be observable directly at PTOLEMY \cite{McKeen:2018xyz,Baracchini:2018wwj}.

Furthermore, since the mechanism studied here is sensitive to the Hubble scale during 
inflation, future measurements of inflationary B-modes at CMB-S4 experiments will have important implications for this class of scenarios. The forecasted sensitivity to the scalar-to-tensor ratio $r \sim 10^{-3}$
implies a sensitivity to $H_I \sim 10^{12}$ GeV \cite{s4collaboration2020cmbs4},
which yields observable $\Delta N_{\rm eff}$ from $V$ decays in the upper half of Fig. \ref{direct}.

Finally, it has been shown that contributions to $\dneff \sim 0.5$ may play an important 
role in alleviating the discrepancy between
early and late time determinations of the  Hubble tension \cite{divalentino2021realm,Knox_2020}.
Although models with nonzero $\dneff$ do not completely eliminate the tension, 
it is intriguing that the contributions required to reduce its statistical signficance are 
readily accommodated in the parameter space studied in this class of models. 

\bigskip

{\it  \noindent Acknowledgments:} 
We thank Asher Berlin, Nikita Blinov,  Dan Hooper, Kevin Kelly, Rocky Kolb, and David McKeen for helpful conversations.  This manuscript has been authored by Fermi Research Alliance, LLC under Contract No. DE-AC02-07CH11359 with the U.S. Department of Energy, Office of High Energy Physics. 

\bibliography{Neff}

\begin{appendix}

\end{appendix}

\end{document}